\documentclass[aps,reprint,pra]{revtex4-1}

\usepackage{amsmath}
\usepackage{dsfont}
\usepackage{graphicx}
\usepackage{hyperref}
\usepackage{dsfont}

\newcommand{\ket}[1]{\left|#1\right>}

\renewcommand{\vec}[1]{\mathbf{#1}}
\newcommand{\Tr}[0]{\operatorname{Tr}}

\begin{document}
 \title{Optimizing electrically controlled echo sequences for the exchange-only qubit}
\author{Niklas Rohling and Guido Burkard}
\affiliation{Department of Physics, University of Konstanz, D-78457 Konstanz, Germany}

\begin{abstract}
Recently, West and Fong [New J.~Phys.~\textbf{14}, 083002 (2012)] introduced an echo scheme
for an exchange-only qubit, which relies entirely on the exchange-interaction.
Here, we compare two different exchange-based sequences and two optimization strategies,
Uhrig dynamical decoupling (UDD)
and optimized filter function dynamical decoupling (OFDD),
which were introduced for a single-spin qubit and are applied
in this paper to the three-spin exchange-only qubit.
The calculation shows that the adaption of the optimization concepts can be achieved by
straight-forward calculation.
We consider two types of noise spectra, Lorentzian and Ohmic noise.
For both spectra, the results reveal a slight dependence of the performance on the
choice of the echo sequence.
\end{abstract}
\maketitle

\section{Introduction}
The concept of exchange-only quantum computation \cite{uqcwtei} relies
on all-electrical qubit control.
Exchange-only qubits are defined as two-dimensional subspaces of three electron spins.
Given the advantages of all-electrical control of these qubits,
it is important to investigate the possible decoherence mechanisms
and their mitigation using appropriate techniques.
While a homogeneous magnetic field of unknown strength
does not harm the qubit state,
an inhomogeneous magnetic field,
which might occur due to nuclear spins in the host material,
can cause decoherence and leakage.
In the case of the resonant-exchange qubit
\cite{PhysRevLett.111.050501,PhysRevLett.111.050502},
the degeneracy of the qubit (and leakage) states is partially lifted
by the always-on exchange coupling.
In an external magnetic field,
the leakage can be suppressed completely and dephasing may be corrected by
an echo sequence resembling the single-spin qubit spin echo\cite{PhysRevB.90.045308}.
For the degenerate exchange-only qubit,
the situation is more complicated due to the existence of a leakage
state \footnote{In a strong external magnetic field only one leakage state
need to be taken into account, see \cite{PhysRevB.90.045308}}.
Applying spin-echo techniques to each spin individually is
not favorable as this requires magnetic control which contradicts
the concept of exchange-only quantum computing \cite{PhysRevB.90.045308}.
West and Fong \cite{West_Fong2012} introduced an echo scheme for the exchange-only qubit
which is based on SWAP operations between neighboring spin states.
These operations are provided directly by the exchange interaction.
The basic idea is to average the acquired phases of the spin states by
permuting their positions within the triple quantum dot.
A sequence which also corrects erroneous SWAP gates was introduced by
Hickman \textit{et al.}~\cite{PhysRevB.88.161303}.

In this paper, we focus on optimization strategies for the timing of the pulses
in exchange-based echo schemes.
We assume that the pulse lengths are negligible and the decoherence occurs between the pulses.
West and Fong \cite{West_Fong2012} already applied
Uhrig dynamical decoupling (UDD) \cite{Uhrig_PRL_2007,UhrigNJP} to their SWAP-based sequence
for the exchange-only qubit.
Here, we transfer the concept of
the optimized filter function dynamical decoupling (OFDD) \cite{uys2009}
from the single-spin to the three-spin system.
Furthermore, we consider two different SWAP sequences and compare their performance
for a simple Carr-Purcell-Meiboom-Gill (CPMG)
timing of the pulse as well as for UDD and OFDD.

The paper is organized as follows.
In Sec.~\ref{sec:system}, a model for the system of interest is introduced.
Sec.~\ref{sec:echo_sequences} contains the calculation to obtain the fidelity in
dependance of the pulse sequence and the noise spectrum.
In Sec.~\ref{sec:optimizing} the optimization strategies for the timing of the pulses
are discussed and the results for the fidelity compared to the unchanged qubit
are presented.
Finally, we conclude in Sec.~\ref{sec:conclusions}.

\section{Exchange-only qubit in a random magnetic field}
\label{sec:system}
\begin{figure}
 \includegraphics[width=\columnwidth]{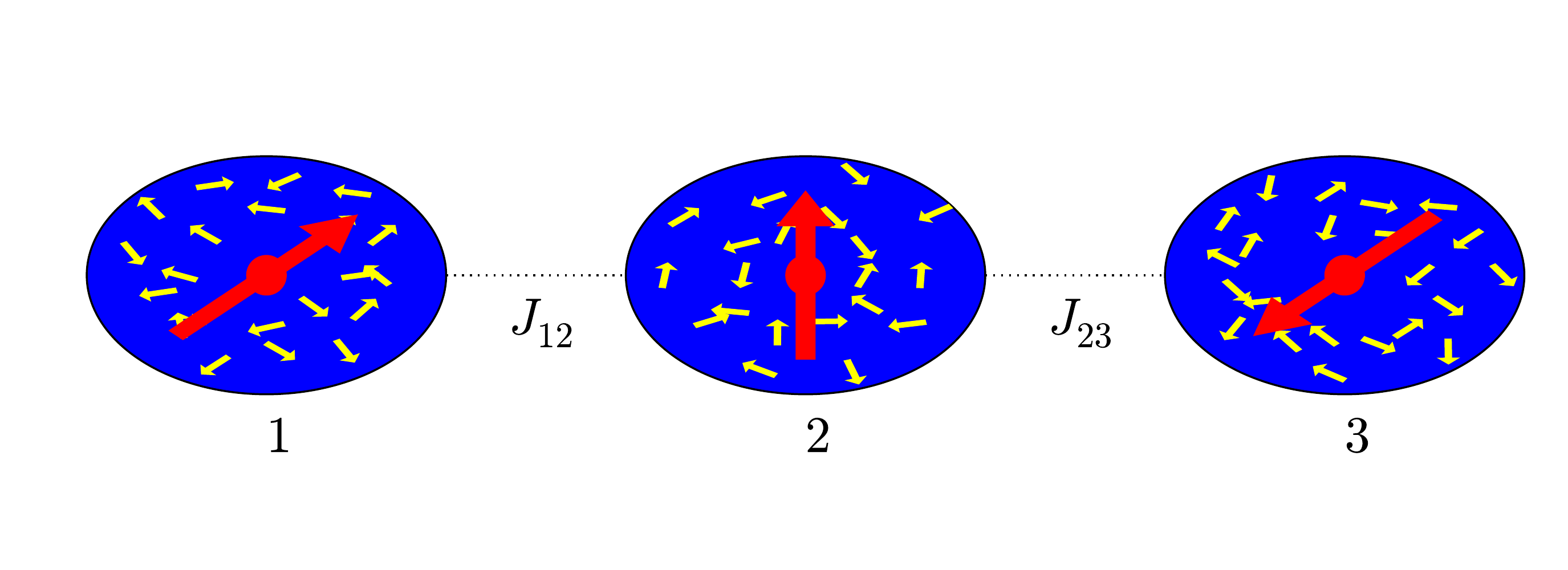}
 \caption{\label{fig:sys}
          (Color online) Sketch of the system of three quantum dots each hosting an electron
          represented by a large red (gray) dot and arrow.
          The electron spin states in dots 1 and 2 (2 and 3) can be coupled via the exchange
          interaction $J_{12}$ ($J_{23}$), while
          there is no direct coupling between dots 1 and 3.
          The electron spins in each dot experience the influence of the nuclear
          spins, which are represented by small yellow (light gray) arrows, via
          the hyperfine coupling.
          We describe this influence of the nuclear spins by fluctuating Overhauser fields.}
\end{figure}

The system which we consider here consists of three quantum dot hosting one
electron spin each, see Fig.~\ref{fig:sys}.
The electron spins are coupled by the exchange interaction
and influenced by local magnetic fields,
\begin{equation}
\label{eq:hamiltonian}
 H = \frac{J_{12}}{4}\boldsymbol\sigma_1\cdot\boldsymbol\sigma_2
     + \frac{J_{23}}{4}\boldsymbol\sigma_2\cdot\boldsymbol\sigma_3
     +\sum_{i=1}^3\vec B_i\cdot\boldsymbol\sigma_i
\end{equation}
where the exchange couplings $J_{12}$ and $J_{23}$ can be electrically
controlled \cite{LoDi1998,petta}.
Here, $\boldsymbol\sigma_i=(\sigma_{ix},\sigma_{iy},\sigma_{iz})^T$
denotes the vector of Pauli operators for the spins in dots $i=1,2,3$.
The magnetic field $\vec B_i=B_\textrm{ext}\vec e_z + \vec B_i^O$ consists of an external
magnetic field $B_\textrm{ext}$ in $z$ direction and the Overhauser field as a
classical model for the interaction with the nuclear spin bath.
In the case $B_\textrm{ext}\gg|\vec B_i^O|$, the dephasing is dominated by the
$z$ component of the Overhauser field under the condition that it is time
independent, see Appendix A.
This still holds for the Overhauser field changing slowly in time compared to the Larmor
precession caused by $B_\textrm{ext}$ within a rotating wave approximation \cite{taylor2006dephasing}.
In the following we consider only the magnetic field in $z$ direction.
Thus, states with different total spin in $z$ direction will not be coupled.
The qubit subspace of the exchange-only qubit is the two-dimensional space
characterized by
the total spin $s=1/2$ and the spin in quantization ($z$) direction $s_z=1/2$, see \cite{uqcwtei}.
Therefore, leakage is possible to the state with $s=3/2$ and $s_z=1/2$
in the presence of a magnetic field gradient in $z$-direction \cite{PhysRevB.90.045308}.

\section{Echo sequences}
\label{sec:echo_sequences}
West and Fong \cite{West_Fong2012} introduced an echo scheme relying only
on the exchange interaction in agreement with the concept of the exchange-only qubit.
They suggested to interchange the spin information of neighboring dots in such a way that
any spin state spends equal time in each of the three dots.
The operations which are needed are SWAP gates for the dots 1 and 2 and for the dots 2 and 3,
${\rm SWAP}_{12}$ and ${\rm SWAP}_{23}$.
These operations are provided directly by the exchange coupling.
It is assumed that this coupling can be tuned to values much larger than
the differences in the Zeeman splitting between the dots.
In this case, the exchange coupling is not disturbed by the Overhauser field and
the gate times can be negligibly short.
West and Fong considered sequences using the gates $P={\rm SWAP}_{23}{\rm SWAP}_{12}$
and $P^{-1}={\rm SWAP}_{12}{\rm SWAP}_{23}$ in alternating pairs,
$P\rightarrow P\rightarrow P^{-1}\rightarrow P^{-1}\rightarrow P\rightarrow P\rightarrow P^{-1}\rightarrow P^{-1}$
and so on.
They showed that the concept of UDD \cite{Uhrig_PRL_2007,UhrigNJP}
removing the influence of the noise up to $m$th order in time ($m=0,1,2,3,\ldots$) can be
applied to this three-spin problem.
In the present paper, we compare the sequence of alternating pairs of
$P$ and $P^{-1}$ to the sequence using only the cyclic permutation of the
spin states, $P$.
Furthermore, we additionally use the concept of optimized noise filtration
\cite{uys2009}.
Both, OFDD and UDD, were originally introduced for a single qubit dephasing
without leakage states.

We now consider the $s_z=+1/2$ subspace, starting from the product basis
$\{\ket{\uparrow\uparrow\downarrow},
   \ket{\uparrow\downarrow\uparrow},
   \ket{\downarrow\uparrow\uparrow}
\}$.
In this basis the term in the Hamiltonian describing the effect of
a time-dependent magnetic field in $z$ direction is diagonal.
The corresponding time evolution for the states
$\ket{\uparrow\uparrow\downarrow}$,
$\ket{\uparrow\downarrow\uparrow}$, and
$\ket{\downarrow\uparrow\uparrow}$
evokes the phase factors,
$e^{-i(\phi_1+\phi_2-\phi_3)}$,
$e^{-i(\phi_1-\phi_2+\phi_3)}$, and
$e^{-i(-\phi_1+\phi_2+\phi_3)}$,
respectively.
Formally, we track the spin state when a SWAP gate transfers it to another quantum  dot.
The time evolution at time $T$ for the spin state which is in the first dot at time $t=0$ is
\begin{equation}
 U_1(T) = e^{-i\phi_1\tilde\sigma_{1z}}~\text{with}~\phi_1=\int\limits_0^Tdt\,h_1(t),
\end{equation}
where $\tilde\sigma_{1z}$ is the Pauli matrix for the individual spin state labeled here
with 1 although the state is stored in dots 2 and 3 for some time.
This spin state experiences the magnetic field $h_1(t)$, which is the field in the
dot where the spin state is stored at time $t$.
For the spin states initially stored in the dots 2 and 3, the time evolution is
given in full analogy by $U_2(T)$ and $U_3(T)$.
We use the states
\begin{equation}
 \ket{\pm} = \frac{\ket{\uparrow\uparrow\downarrow}
                   + e^{\pm i2\pi/3}\ket{\uparrow\downarrow\uparrow}
                   + e^{\mp i2\pi/3}\ket{\downarrow\uparrow\uparrow}}
                  {\sqrt{3}}
\end{equation}
as an orthogonal basis of the qubit subspace.
In the corresponding Bloch sphere, with the poles $\ket{\pm}$,
the eigenstates of the exchange interactions between neighboring dots
lie in the equatorial plane.
Therefore, ${\rm SWAP}_{23}$ and ${\rm SWAP}_{12}$ interchange $\ket{+}$ and $\ket{-}$,
i.e.,
${\rm SWAP}_{12}\ket{\pm}=\ket{\mp}$ and ${\rm SWAP}_{23}\ket{\pm}=\ket{\mp}$,
while $\ket{+}$ and $\ket{-}$ are left unchanged by $P$ and $P^{-1}$.
The only relevant leakage state is
\begin{equation}
  \ket{L} = \frac{\ket{\uparrow\uparrow\downarrow}
                   + \ket{\uparrow\downarrow\uparrow}
                   + \ket{\downarrow\uparrow\uparrow}}
                  {\sqrt{3}}
\end{equation}
with the quantum numbers $s=3/2$ and $s_z=1/2$.
In the basis $\{\ket{+},\ket{-},\ket{L}\}$
the time evolution is represented by the matrix
\begin{widetext}
\begin{equation}
\label{eq:matrix}
U(T) = \frac{1}{3}
        \left(
        \begin{array}{ccc}
           e^{i\varphi_3}{+}e^{i\varphi_2}{+}e^{i\varphi_1}
         & e^{i\varphi_3}{+}e^{i(\varphi_2+\frac{2\pi}{3})}{+}e^{i(\varphi_1-\frac{2\pi}{3})}
         & e^{i\varphi_3}{+}e^{i(\varphi_2-\frac{2\pi}{3})}{+}e^{i(\varphi_1+\frac{2\pi}{3})} \\
           e^{i\varphi_3}{+}e^{i(\varphi_2-\frac{2\pi}{3})}{+}e^{i(\varphi_1+\frac{2\pi}{3})}
         & e^{i\varphi_3}{+}e^{i\varphi_2}{+}e^{i\varphi_1}
         & e^{i\varphi_3}{+}e^{i(\varphi_2+\frac{2\pi}{3})}{+}e^{i(\varphi_1-\frac{2\pi}{3})}\\
           e^{i\varphi_3}{+}e^{i(\varphi_2+\frac{2\pi}{3})}{+}e^{i(\varphi_1-\frac{2\pi}{3})}
         & e^{i\varphi_3}{+}e^{i(\varphi_2-\frac{2\pi}{3})}{+}e^{i(\varphi_1+\frac{2\pi}{3})}
         & e^{i\varphi_3}{+}e^{i\varphi_2}{+}e^{i\varphi_1}
          \end{array}
          \right)
\end{equation}
\end{widetext}
with $\varphi_j=2\phi_j-\phi_1-\phi_2-\phi_3$.
In this matrix, we denote the upper left $2\times2$ block, which describes the dynamics within the qubit subspace, by $U_Q(T)$.
We compare this operation with the perfect storage of the qubit, $U_Q(T)=\mathds{1}$.
In this context, $T$ denotes the desired storage time of the qubit.
The fidelity of the time evolution with respect to the identity operation is \cite{Pedersen200747}
\begin{equation}
 F = \frac{\Tr(U_Q^\dagger U_Q)+|\Tr(U_Q)|^2}{6}.
\end{equation}
Note that $U_Q$ is not unitary.
From Eq.~(\ref{eq:matrix}), we obtain
\begin{equation}
\begin{split}
\label{eq:fid_simp}
 F ={} &\frac{4}{9}+\sum_{i<j}\left\{\frac{2\cos(2[\phi_i{-}\phi_j])}{9}\right.\\
     & \left.+\frac{\cos(2[\phi_i{-}\phi_j]{+}\tfrac{2\pi}{3})+\cos(2[\phi_i{-}\phi_j]{-}\tfrac{2\pi}{3})}{27}\right\}.
\end{split}
\end{equation}
We assume that the Overhauser fields can be described by a Gaussian distribution.
Then the same holds for the acquired phases, thus we obtain
\begin{equation}
 \begin{split}
  \langle\cos(2(\phi_1-\phi_2))\rangle & = e^{-2\langle(\phi_1-\phi_2)^2\rangle},\\
  \langle\cos(2(\phi_1-\phi_2)\pm\tfrac{2\pi}{3})\rangle & = -\frac{1}{2}e^{-2\langle(\phi_1-\phi_2)^2\rangle},
 \end{split}
\end{equation}
in analogy with the case of a single spin \cite{UhrigNJP}.
In order to calculate $\langle(\phi_1-\phi_2)^2\rangle$, West and Fong \cite{West_Fong2012}
introduced the functions $f_j(t)$, $j=1,2,3$, which are defined according to the position of
the spin states: For the initial positions, (1,2,3), where the numbers 1, 2, and 3 are the
labels of the spin states, 
the values of the functions are $\{f_1,f_2,f_3\}=\{1,-1,0\}$.
For the positions (2,3,1), the functions are $\{f_1,f_2,f_3\}=\{-1,0,1\}$
and for the positions (3,2,1), they are $\{f_1,f_2,f_3\}=\{0,1,-1\}$.
The Overhauser fields are labeled according to the quantum dot where they can be found
by $B_j(t)$ for dot number $j$, see Eq.~(\ref{eq:hamiltonian}).
Then the phase difference between two spin states at time $T$ is given by \cite{West_Fong2012}
\begin{equation}
\begin{split}
 \phi_1(T){-}\phi_2(T)& = \int\limits_0^T\!dt\,[h_1(t)-h_2(t)]\\
                    &\!\! = \int\limits_0^T\!dt\,[f_1(t)B_1(t){+}f_2(t)B_2(t){+}f_3(t)B_3(t)].
\end{split}
\end{equation}
The expressions for $\phi_2-\phi_3$ and $\phi_3-\phi_1$ can be obtained by permuting
the indices of the functions $f_j(t)$.
The variance is
\begin{equation}
\begin{split}
 & \langle(\phi_1(T)-\phi_2(T))^2\rangle \\
  =& \left\langle \int\limits_0^Tdt_1\,[f_1(t_1)B_1(t_1){+}f_2(t_1)B_2(t_1){+}f_3(t_1)B_3(t_1)]\right.\\
 & ~~\left. \times\int\limits_0^Tdt_2\,[f_1(t_2)B_1(t_2){+}f_2(t_2)B_2(t_2){+}f_3(t_2)B_3(t_2)]\right\rangle\\
  =&  \sum_{i,j\in\{1,2,3\}}\int\limits_0^Tdt_1\,\int\limits_0^Tdt_2f_i(t_1)f_j(t_2)
       \langle B_i(t_1)B_j(t_2)\rangle\\
  = & \frac{1}{\pi}\sum_{i,j\in\{1,2,3\}}\int\limits_0^\infty d\omega\, y_i(\omega T)y_j^*(\omega T)\frac{p_{ij}(\omega)}{\omega^2}
 \end{split}
\end{equation}
with $y_j(\omega T):=\frac{\omega}{i}\int_0^T dt\,e^{i\omega t}f_j(t)$ being the filter
function, which equals, up to the factor of $\omega/i$, the Fourier transform of
the switching function $f_j(t)$, $j=1,2,3$.
The function $p_{ij}(\omega)$ is the power spectrum of $\langle B_i(t)B_j(0)\rangle$,
$ p_{ij}(\omega) = 2\int_0^\infty dt\,\cos(\omega t)\langle B_i(t)B_j(0)\rangle$.
Under the assumption that the Overhauser fields in the dots have the same variance
and the same power spectrum while being independent of each other,
$p_{ij}(\omega)=\delta_{ij}p(\omega)$,
the variance of the phase differences can be written as
\begin{equation}
\label{eq:variance}
\begin{split}
   & \langle(\phi_1(T)-\phi_2(T))^2\rangle\\
 = & \langle(\phi_2(T)-\phi_3(T))^2\rangle\\
 = & \langle(\phi_3(T)-\phi_1(T))^2\rangle\\
 = & \frac{1}{\pi}\int\limits_0^\infty d\omega\,
 \underbrace{[|y_1(\omega)|^2 + |y_2(\omega)|^2 + |y_3(\omega)|^2]}_{=F_F(\omega T)}\frac{p(\omega)}{\omega^2}.
 \end{split}
\end{equation}
Here $F_F(\omega T)$ is the filter function of the SWAP-based echo sequence
which determines the values of $f_j(t)$ at times $t\in(0,T)$.
The assumption that the Overhauser fields are described by the same random distribution
also lead to a further simplification of the expression Eq.~(\ref{eq:fid_simp}) for the fidelity,
which assumes the form
\begin{equation}
 F = \frac{4}{9}+\frac{5}{9}e^{-2\langle(\phi_1-\phi_2)^2\rangle}.
\end{equation}
The value for $\langle(\phi_1-\phi_2)^2\rangle$ will depend on the
noise spectrum $p(\omega)$, the pulse sequence, and the time $T$.
Here we focus on Ohmic noise,
\begin{equation}
\label{eq:ohm}
 p_\text{Ohm}(\omega)=\omega\Theta(\omega_1-\omega),
\end{equation}
with a cutoff described by the Heaviside function $\Theta(\cdot)$, and Lorentzian noise,
\begin{equation}
\label{eq:lorentz}
 p_\text{Lorentz}(\omega)=\frac{\omega_1}{1+(\frac{\omega}{\omega_1})^2}.
\end{equation}
The latter has been used for a model to explain experiments with a nuclear spin bath \cite{de_Lange2010}.
In the situation considered in this paper, the random field is also assumed to originate
from the nuclear spins.
Nevertheless, we perform the calculations for the Ohmic noise spectrum as well
to demonstrate that the method can be applied to different noise spectra.
Note that the parameter $\omega_1$
is a sharp cutoff in Eq.~(\ref{eq:ohm}) while it is a parameter determining
the width of the spectrum in  Eq.~(\ref{eq:lorentz}).
Here we are interested in the scaling behavior of the fidelity with respect
to the parameter $\omega_1$.
Therefore, it is possible to set the relative noise strength to one,
see Ref.~\cite{uys2009}.
The generalization to an arbitrary noise strength is straightforward.
As done by Uys \textit{et al.}~in Ref.~\cite{uys2009} we express the integral in Eq.~(\ref{eq:variance}) by
using the dimensionless variables $T'=T\omega_1$ and $\omega'=\omega/\omega_1$.
Then the variance of the phase difference reads
\begin{equation}
\label{eq:phase_diff_int}
 \langle(\phi_1-\phi_2)^2\rangle = \int\limits_0^\infty d\omega'\,F_F(\omega'T')\frac{\tilde p(\omega')}{{\omega'}^2},
\end{equation}
where the noise spectrum is rescaled,
\begin{equation}
 \tilde p(\omega')=\tilde p_\textrm{Ohm}(\omega')=\omega'\Theta(1-\omega')
\end{equation}
or
\begin{equation}
 \tilde p(\omega')=\tilde p_\text{Lorentz}(\omega')=\frac{1}{1+{\omega'}^2}.
\end{equation}
The filter function depends on the sequence of applied SWAP operations
via the switching functions $f_j(t)$, $j=1,2,3$.
First we consider the same operation $P=\text{SWAP}_{23}\text{SWAP}_{12}$ applied
at the times $T\delta_j$, $j=1,\ldots,n$.
Then the switching functions are periodic with respect to the time intervals $[\delta_j,\delta_{j{+}1})$
with period three.
The switching functions $f_1$ and $f_2$ at time $t$ are given by
\begin{equation}
 \{f_1,f_2\}=\begin{cases}
                           \{1,-1\} & \text{if}~t/T\in[\delta_{j},\delta_{j+1}),~j\!\!\!\!\mod3=0,\\
                           \{-1,0\} & \text{if}~t/T\in[\delta_{j},\delta_{j+1}),~j\!\!\!\!\mod3=1,\\
                           \{0,1\} & \text{if}~t/T\in[\delta_{j},\delta_{j+1}),~j\!\!\!\!\mod3=2.
                          \end{cases}
\end{equation}
The third function is always determined by $f_3(t)=-(f_1(t)+f_2(t))$.
For convenience, $\delta_0=0$ and $\delta_{n+1}=1$ have been introduced.
For the sequence with alternating pairs of $P$ and $P^{-1}$, which was considered in Ref.~\cite{West_Fong2012},
the switching functions at time $t$ are
\begin{equation}
 \{f_1,f_2\}=\begin{cases}
                           \{1,-1\} & \!\text{if}~t/T\in[\delta_{j},\delta_{j+1}),~j\!\!\!\!\mod4{=}0,\\
                           \{{-}1,0\} & \!\text{if}~t/T\in[\delta_{j},\delta_{j+1}),~j\!\!\!\!\mod4{=}1\,\text{or}\,3,\\
                           \{0,1\} & \!\text{if}~t/T\in[\delta_{j},\delta_{j+1}),~j\!\!\!\!\mod4{=}2,
                          \end{cases}
\end{equation}
and again $f_3(t)=-(f_1(t)+f_2(t))$.
We compare the results for the exchange based SWAP sequences to individual spin echoes,
where a $\sigma_x$ gate is applied on each spin at times $\delta_jT$, $j=1,\ldots,n$.
This single spin manipulation is not compatible with the concept of exchange-only
quantum computing as it requires single-spin manipulation with a time-dependent local magnetic field.
It is considered here for comparison of the efficiency of the echo sequences only.

\section{Waiting time optimization strategies}
\label{sec:optimizing}
In this section we apply different concepts for optimizing $\{\delta_1,\ldots,\delta_n\}$.
These concepts have been introduced for single-spin echoes but can be applied for the
three spin system as well.
The calculation of the filter function is straight forward for a given sequence and
$\langle(\phi_1-\phi_2)^2\rangle$ can be calculated with Eq.~(\ref{eq:phase_diff_int})
by solving the respective integrals.

\subsection{CPMG sequence}
\begin{figure}
 \includegraphics{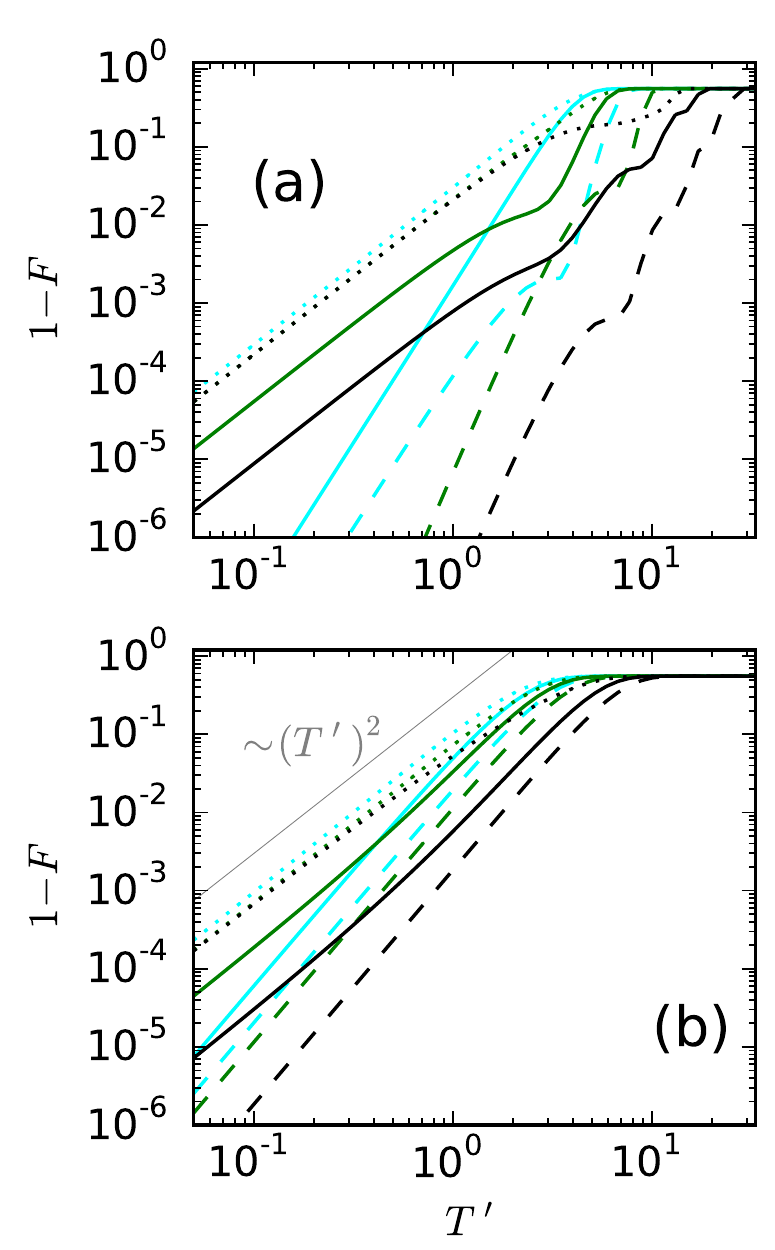}
\caption{(Color online)
Infidelity $1-F$ for Ohmic noise (a) and Lorentzian noise (b)
in dependence of dimensionless storage time $T'$ 
for pulse times chosen according to the CPMG scheme.
The number of applied pulses $n$ are $n=3$ (cyan, light gray), $n=4$ (green, dark gray), and $n=10$ (black).
The applied sequences are the all-cyclic permutations realized by applying
$P={\rm SWAP}_{12}{\rm SWAP}_{23}$ at every time $T\delta_j$ (solid lines),
the sequence of pairs of $P$ and $P^{-1}$ (dotted lines) and the single spin
operations (dashed lines).
}
\label{fig:cpmg}
\end{figure}
The CPMG sequence \cite{CarrPurcell, Meiboom_Gill} is defined by
$\delta_1=1/(2n)$, $\delta_j=\delta_{j-1}+1/n$ for $j=2,\ldots,n$.
The waiting times $t_w$ between consecutive pulses are always the same.
The waiting time between initialization and the first pulse equals the waiting time between the last pulse
and the measurement at time $T$ and is half as long as $t_w$.
The infidelity $1-F$ for this chosen timing is presented in Fig.~\ref{fig:cpmg}.
The fidelity $F$ can be increased with an increasing number of pulses.
This strategy is more effective for the Ohmic noise, which is stronger at higher
frequencies compared to the Lorentzian noise.
Note that the better scaling behavior of the infidelity in Fig.~\ref{fig:cpmg}
for the all cyclic sequence and $n=3$ ($P\rightarrow P\rightarrow P$)
originates from the fact that the filter function vanishes up to first order
for this sequence if $n$ is an integer multiple of 3.

\subsection{Applying Uhrig-type dynamical decoupling}
\begin{figure}
 \includegraphics{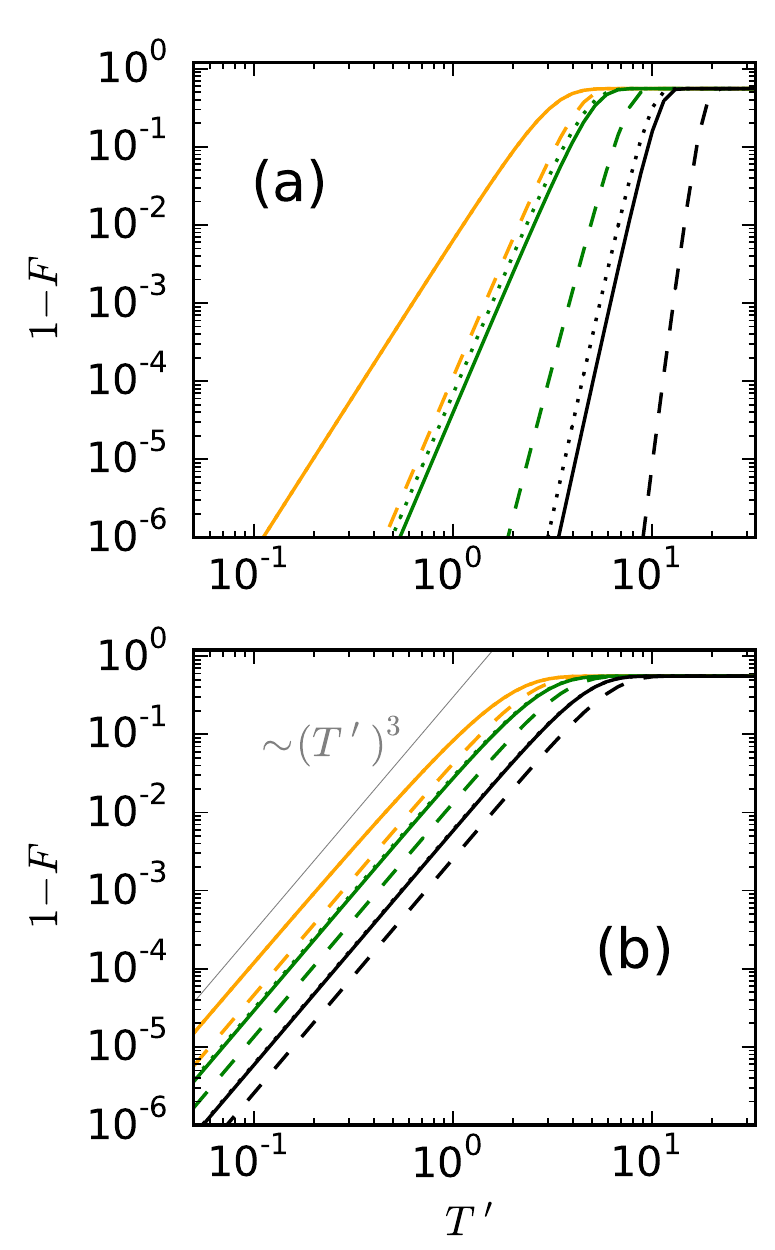}
\caption{(Color online)
Infidelity $1-F$ in dependence of dimensionless storage time $T'$ for Ohmic noise (a)
and Lorentzian noise (b) for pulse times chosen according to the UDD optimization strategy.
The number of applied pulses $n$ are $n=2$ (orange, light gray), $n=4$ (green, dark gray), and $n=10$ (black).
The applied sequences are plotted in the same line styles as in Fig.~\ref{fig:cpmg}.
In (b) the results for the exchange-based pulses with the all-cyclic permutations
and the pairs of $P$ and $P^{-1}$ are very similar,
thus the solid and the dotted lines are on top of each other.
For $n=2$ these sequences are identical by definition.
The better performance of the single-spin pulses (dashed lines) compared to the
SWAP-based sequences for the same number of pulses, $n$, is due to the fact that
it allows for the filter function to be zero up to the order $n$ while it is only order $n/2$
for the SWAP-based pulse sequences.
}
\label{fig:udd}
\end{figure}

The concept of UDD \cite{Uhrig_PRL_2007,UhrigNJP} relies on the
requirement that the filter function should be zero up to an order $m$,
\begin{equation}
 \left.\left(\frac{\partial}{\partial(\omega'T')}\right)^kF_F(\omega'T')\right|_{\omega'T'=0}=0,~~ k=0,\ldots,m.
\end{equation}
Uhrig showed that this can be achieved in the case of a single spin by $n=m$ pulses \cite{Uhrig_PRL_2007}.
Moreover, the values for $\delta_j$ are given by the analytical expression
$\delta_j=\{1 + \sin(\pi j/[n + 1])\}/2$ \cite{Uhrig_PRL_2007} in this single-spin case.
West and Fong \cite{West_Fong2012} extended this concept to the exchange-only qubit
for the sequence $P\to P\to P^{-1}\to P^{-1}\ldots$.
In this situation the number of pulses has to be $n=2m$.
Here, we also apply the concept to the sequence $P\to P\to P\ldots$ again with $n=2m$.
The values of $\delta_j$ are different from the values in the West-Fong sequence for $n>2$.
For $n=2$ the two sequences are identical with $\delta_1=1/3$ and $\delta_2=2/3$.
In Fig.~\ref{fig:udd}, we also present the infidelities for $n=4$ with
$\delta_1=1-\delta_4=1/6$,
$\delta_2=1-\delta_3=1/3$,
and for $n=10$ with
$\delta_1=0.0422244245173296$,
$\delta_2=0.0940587956886883$,
$\delta_3=0.2172228408817372$,
$\delta_4=0.2838895075484039$,
$\delta_5=0.4518343711713587$,
and $\delta_j=1-\delta_{11-j}$.
The numbers are different from the solutions for the West-Fong sequence \cite{West_Fong2012}.
In general, the values of $\delta_j$ can be found numerically while a closed expression is unknown
for the SWAP-based sequences, see also \cite{West_Fong2012}.
Comparing Fig.~\ref{fig:udd} to Fig.~\ref{fig:cpmg}, we see that UDD can outperform the CPMG
sequence where the improvement is more significant for the Ohmic noise than for the Lorentzian noise.

\subsection{Applying optimized noise filtration}
\begin{figure}
 \includegraphics{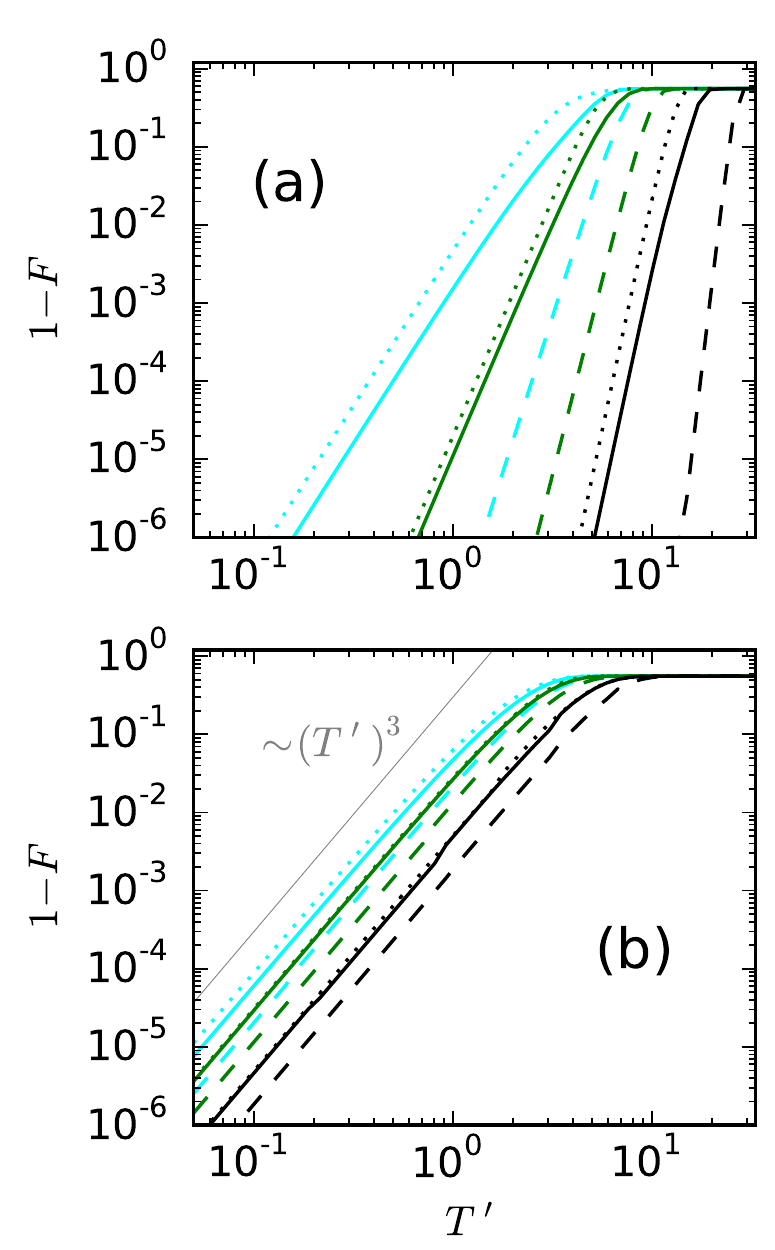}
\caption{(Color online)
Infidelity $1-F$ in dependence of dimensionless storage time $T'$ for Ohmic noise (a)
and Lorentzian noise (b) for pulse times chosen according to OFDD.
The color codes and the line styles are identical to Fig.~\ref{fig:cpmg}.
}
\label{fig:ofdd}
\end{figure}

Another strategy for minimizing the dephasing,
OFDD, was introduced
by Uys \textit{et al.}~\cite{uys2009}.
In this concept, the integral
\begin{equation}
 \label{eq:ofdd_condition}
 \int\limits_0^1d\omega'\,F_F(\omega'T')
\end{equation}
is minimized by finding a suitable set of
$\{\delta_1,\ldots,\delta_n\}$.
In Ref.~\cite{uys2009} OFDD was explicitly considered for a single spin.
Here we apply the method for the three-spin problem.
For the considered sequences, the integrals included in (\ref{eq:ofdd_condition})
can be treated analytically.
The results for the infidelity are shown in Fig.~\ref{fig:ofdd}.
Similar to the results for the single spin \cite{uys2009},
we find that OFDD can lead to improved fidelities compared to UDD.

\section{Conclusions}
\label{sec:conclusions}
In this paper, we have considered exchange-only based echo sequences
for three electron spins.
We have shown that, in addition to the UDD-like optimized sequences
introduced by West and Fong \cite{West_Fong2012},
the concept of OFDD can be applied to this three-spin case as well.
We compared two different sequences of SWAP-based echo sequences,
the one used in Ref.~\cite{West_Fong2012} and one which applies the
same operation at every time $\delta_jT$.
The optimal times according to UDD and OFDD depend on the choice of the sequence.
The fidelity depends slightly on this choice with a small advantage for
the all-cyclic permutation.
The improvement of the fidelity by the optimization strategies UDD and OFDD
compared to the CPMG sequence is more significant for Ohmic noise than for
Lorentzian noise.
This could have been expected with respect to results for a single spin under
the influence of high- and low-frequency noise \cite{uys2009}.

\begin{acknowledgments}
We thank Maximilian Russ for usefull discussions
and the DFG for financial support under the
programs SPP 1285 and SFB 767.
GB acknowledges financial support from ARO
through grant No.~W911NF-15-1-0149.
\end{acknowledgments}

\section*{Appendix A: Fidelity of a spin 1/2 in a quasi static magnetic field}
The fidelity of a quantum gate $U$ acting on a single qubit in comparison
to a desired gate $U_0$ is given by \cite{Pedersen200747}
\begin{equation}
F = \frac{2 + |\operatorname{Tr}(U_0U^\dagger)|^2}{6}.
 \tag{A1}
\end{equation}
Here we define $U_0$ as the single-qubit rotation around the $z$ axis
about the angle $B_{\rm ext}t$, i.e. the time evolution operator at
time $t$ for $\vec B^O=0$.
The operator $U$ is the time evolution under the Hamiltonian
$H = (B_{\rm ext}\vec e_z + \vec B^O)\cdot\boldsymbol\sigma$
with an unknown, in general non-zero Overhauser field $\vec B^O$.
We find at time $t$
\begin{equation}
\begin{split}
 & \operatorname{Tr}(U_0U^\dagger) =  2\cos((\omega{-}B_{\rm ext})t)\\
 & -(\cos((\omega{-}B_{\rm ext})t)-\cos((\omega{+}B_{\rm ext})t))(1-\tfrac{B_{\rm ext}{+}B_z^O}{\omega})
  \end{split}
 \tag{A2}
 \label{eq:trace}
\end{equation}
with $\omega=\sqrt{(B_{\rm ext}{+}B_z^O)^2 + (B_x^O)^2 + (B_y^O)^2}$.
For $B_{\rm ext}\gg |\vec B^O|$ in lowest order,
\begin{equation}
1 - \frac{B_{\rm ext}{+}B_z^O}{\omega} \approx \frac{(B_x^O)^2 + (B_y^O)^2}{(B^{\rm ext})^2}
 \tag{A3}
\end{equation}
which is negligible.
Therefore the respective term in Eq.~(\ref{eq:trace}) is small and can
be neglected for all times $t$.
The fidelity $F$ is therefore
approximated by
\begin{equation}
 F \approx \frac{2 + 4\cos^2(B_z^Ot)}{6}
 \tag{A4}
\end{equation}
for times $t\lesssim 2\pi/(\omega-B_{\rm ext})$
where we expanded $\omega$ in linear order in $\vec B^O$.
This means that on relevant time scales, i.e. where the gate fidelity $F$ is
still close to one, the dephasing due to an Overhauser field is
determined by the component parallel to the (strong) external magnetic field.
Therefore, we have to deal only with commuting operators within the echo schemes
discussed in this paper.


\begin{thebibliography}{17}%
\makeatletter
\providecommand \@ifxundefined [1]{%
 \@ifx{#1\undefined}
}%
\providecommand \@ifnum [1]{%
 \ifnum #1\expandafter \@firstoftwo
 \else \expandafter \@secondoftwo
 \fi
}%
\providecommand \@ifx [1]{%
 \ifx #1\expandafter \@firstoftwo
 \else \expandafter \@secondoftwo
 \fi
}%
\providecommand \natexlab [1]{#1}%
\providecommand \enquote  [1]{``#1''}%
\providecommand \bibnamefont  [1]{#1}%
\providecommand \bibfnamefont [1]{#1}%
\providecommand \citenamefont [1]{#1}%
\providecommand \href@noop [0]{\@secondoftwo}%
\providecommand \href [0]{\begingroup \@sanitize@url \@href}%
\providecommand \@href[1]{\@@startlink{#1}\@@href}%
\providecommand \@@href[1]{\endgroup#1\@@endlink}%
\providecommand \@sanitize@url [0]{\catcode `\\12\catcode `\$12\catcode
  `\&12\catcode `\#12\catcode `\^12\catcode `\_12\catcode `\%12\relax}%
\providecommand \@@startlink[1]{}%
\providecommand \@@endlink[0]{}%
\providecommand \url  [0]{\begingroup\@sanitize@url \@url }%
\providecommand \@url [1]{\endgroup\@href {#1}{\urlprefix }}%
\providecommand \urlprefix  [0]{URL }%
\providecommand \Eprint [0]{\href }%
\providecommand \doibase [0]{http://dx.doi.org/}%
\providecommand \selectlanguage [0]{\@gobble}%
\providecommand \bibinfo  [0]{\@secondoftwo}%
\providecommand \bibfield  [0]{\@secondoftwo}%
\providecommand \translation [1]{[#1]}%
\providecommand \BibitemOpen [0]{}%
\providecommand \bibitemStop [0]{}%
\providecommand \bibitemNoStop [0]{.\EOS\space}%
\providecommand \EOS [0]{\spacefactor3000\relax}%
\providecommand \BibitemShut  [1]{\csname bibitem#1\endcsname}%
\let\auto@bib@innerbib\@empty
\bibitem{uqcwtei}%
  \BibitemOpen
  \bibfield  {author} {\bibinfo {author} {\bibfnamefont {D.~P.}\ \bibnamefont
  {DiVincenzo}}, \bibinfo {author} {\bibfnamefont {D.}~\bibnamefont {Bacon}},
  \bibinfo {author} {\bibfnamefont {J.}~\bibnamefont {Kempe}}, \bibinfo
  {author} {\bibfnamefont {G.}~\bibnamefont {Burkard}}, \ and\ \bibinfo
  {author} {\bibfnamefont {K.~B.}\ \bibnamefont {Whaley}},\ }\href {\doibase
  10.1038/35042541} {\bibfield  {journal} {\bibinfo  {journal} {Nature
  (London)}\ }\textbf {\bibinfo {volume} {408}},\ \bibinfo {pages} {339}
  (\bibinfo {year} {2000})}\BibitemShut {NoStop}%
\bibitem {PhysRevLett.111.050501}%
  \BibitemOpen
  \bibfield  {author} {\bibinfo {author} {\bibfnamefont {J.}~\bibnamefont
  {Medford}}, \bibinfo {author} {\bibfnamefont {J.}~\bibnamefont {Beil}},
  \bibinfo {author} {\bibfnamefont {J.~M.}\ \bibnamefont {Taylor}}, \bibinfo
  {author} {\bibfnamefont {E.~I.}\ \bibnamefont {Rashba}}, \bibinfo {author}
  {\bibfnamefont {H.}~\bibnamefont {Lu}}, \bibinfo {author} {\bibfnamefont
  {A.~C.}\ \bibnamefont {Gossard}}, \ and\ \bibinfo {author} {\bibfnamefont
  {C.~M.}\ \bibnamefont {Marcus}},\ }\href {\doibase
  10.1103/PhysRevLett.111.050501} {\bibfield  {journal} {\bibinfo  {journal}
  {Phys. Rev. Lett.}\ }\textbf {\bibinfo {volume} {111}},\ \bibinfo {pages}
  {050501} (\bibinfo {year} {2013})}\BibitemShut {NoStop}%
\bibitem {PhysRevLett.111.050502}%
  \BibitemOpen
  \bibfield  {author} {\bibinfo {author} {\bibfnamefont {J.~M.}\ \bibnamefont
  {Taylor}}, \bibinfo {author} {\bibfnamefont {V.}~\bibnamefont {Srinivasa}}, \
  and\ \bibinfo {author} {\bibfnamefont {J.}~\bibnamefont {Medford}},\ }\href
  {\doibase 10.1103/PhysRevLett.111.050502} {\bibfield  {journal} {\bibinfo
  {journal} {Phys. Rev. Lett.}\ }\textbf {\bibinfo {volume} {111}},\ \bibinfo
  {pages} {050502} (\bibinfo {year} {2013})}\BibitemShut {NoStop}%
\bibitem {PhysRevB.90.045308}%
  \BibitemOpen
  \bibfield  {author} {\bibinfo {author} {\bibfnamefont {J.-T.}\ \bibnamefont
  {Hung}}, \bibinfo {author} {\bibfnamefont {J.}~\bibnamefont {Fei}}, \bibinfo
  {author} {\bibfnamefont {M.}~\bibnamefont {Friesen}}, \ and\ \bibinfo
  {author} {\bibfnamefont {X.}~\bibnamefont {Hu}},\ }\href {\doibase
  10.1103/PhysRevB.90.045308} {\bibfield  {journal} {\bibinfo  {journal} {Phys.
  Rev. B}\ }\textbf {\bibinfo {volume} {90}},\ \bibinfo {pages} {045308}
  (\bibinfo {year} {2014})}\BibitemShut {NoStop}%
\bibitem {Note1}%
  \BibitemOpen
  \bibinfo {note} {In a strong external magnetic field only one leakage state
  need to be taken into account, see \cite {PhysRevB.90.045308}}\BibitemShut
  {NoStop}%
\bibitem {West_Fong2012}%
  \BibitemOpen
  \bibfield  {author} {\bibinfo {author} {\bibfnamefont {J.~R.}\ \bibnamefont
  {West}}\ and\ \bibinfo {author} {\bibfnamefont {B.~H.}\ \bibnamefont
  {Fong}},\ }\href {http://stacks.iop.org/1367-2630/14/i=8/a=083002} {\bibfield
   {journal} {\bibinfo  {journal} {New Journal of Physics}\ }\textbf {\bibinfo
  {volume} {14}},\ \bibinfo {pages} {083002} (\bibinfo {year}
  {2012})}\BibitemShut {NoStop}%
\bibitem {PhysRevB.88.161303}%
  \BibitemOpen
  \bibfield  {author} {\bibinfo {author} {\bibfnamefont {G.~T.}\ \bibnamefont
  {Hickman}}, \bibinfo {author} {\bibfnamefont {X.}~\bibnamefont {Wang}},
  \bibinfo {author} {\bibfnamefont {J.~P.}\ \bibnamefont {Kestner}}, \ and\
  \bibinfo {author} {\bibfnamefont {S.}~\bibnamefont {Das~Sarma}},\ }\href
  {\doibase 10.1103/PhysRevB.88.161303} {\bibfield  {journal} {\bibinfo
  {journal} {Phys. Rev. B}\ }\textbf {\bibinfo {volume} {88}},\ \bibinfo
  {pages} {161303} (\bibinfo {year} {2013})}\BibitemShut {NoStop}%
\bibitem {Uhrig_PRL_2007}%
  \BibitemOpen
  \bibfield  {author} {\bibinfo {author} {\bibfnamefont {G.~S.}\ \bibnamefont
  {Uhrig}},\ }\href {\doibase 10.1103/PhysRevLett.98.100504} {\bibfield
  {journal} {\bibinfo  {journal} {Phys. Rev. Lett.}\ }\textbf {\bibinfo
  {volume} {98}},\ \bibinfo {pages} {100504} (\bibinfo {year}
  {2007})}\BibitemShut {NoStop}%
\bibitem {UhrigNJP}%
  \BibitemOpen
  \bibfield  {author} {\bibinfo {author} {\bibfnamefont {G.~S.}\ \bibnamefont
  {Uhrig}},\ }\href {http://stacks.iop.org/1367-2630/10/i=8/a=083024}
  {\bibfield  {journal} {\bibinfo  {journal} {New Journal of Physics}\ }\textbf
  {\bibinfo {volume} {10}},\ \bibinfo {pages} {083024} (\bibinfo {year}
  {2008})}\BibitemShut {NoStop}%
\bibitem {uys2009}%
  \BibitemOpen
  \bibfield  {author} {\bibinfo {author} {\bibfnamefont {H.}~\bibnamefont
  {Uys}}, \bibinfo {author} {\bibfnamefont {M.~J.}\ \bibnamefont {Biercuk}}, \
  and\ \bibinfo {author} {\bibfnamefont {J.~J.}\ \bibnamefont {Bollinger}},\
  }\href {\doibase 10.1103/PhysRevLett.103.040501} {\bibfield  {journal}
  {\bibinfo  {journal} {Phys. Rev. Lett.}\ }\textbf {\bibinfo {volume} {103}},\
  \bibinfo {pages} {040501} (\bibinfo {year} {2009})}\BibitemShut {NoStop}%
\bibitem {LoDi1998}%
  \BibitemOpen
  \bibfield  {author} {\bibinfo {author} {\bibfnamefont {D.}~\bibnamefont
  {Loss}}\ and\ \bibinfo {author} {\bibfnamefont {D.~P.}\ \bibnamefont
  {DiVincenzo}},\ }\href {\doibase 10.1103/PhysRevA.57.120} {\bibfield
  {journal} {\bibinfo  {journal} {Phys. Rev. A}\ }\textbf {\bibinfo {volume}
  {57}},\ \bibinfo {pages} {120} (\bibinfo {year} {1998})}\BibitemShut
  {NoStop}%
\bibitem {petta}%
  \BibitemOpen
  \bibfield  {author} {\bibinfo {author} {\bibfnamefont {J.~R.}\ \bibnamefont
  {Petta}}, \bibinfo {author} {\bibfnamefont {A.~C.}\ \bibnamefont {Johnson}},
  \bibinfo {author} {\bibfnamefont {J.~M.}\ \bibnamefont {Taylor}}, \bibinfo
  {author} {\bibfnamefont {E.~A.}\ \bibnamefont {Laird}}, \bibinfo {author}
  {\bibfnamefont {A.}~\bibnamefont {Yacoby}}, \bibinfo {author} {\bibfnamefont
  {M.~D.}\ \bibnamefont {Lukin}}, \bibinfo {author} {\bibfnamefont {C.~M.}\
  \bibnamefont {Marcus}}, \bibinfo {author} {\bibfnamefont {M.~P.}\
  \bibnamefont {Hanson}}, \ and\ \bibinfo {author} {\bibfnamefont {A.~C.}\
  \bibnamefont {Gossard}},\ }\href {\doibase 10.1126/science.1116955}
  {\bibfield  {journal} {\bibinfo  {journal} {Science}\ }\textbf {\bibinfo
  {volume} {309}},\ \bibinfo {pages} {2180} (\bibinfo {year}
  {2005})}\BibitemShut {NoStop}%
\bibitem {taylor2006dephasing}%
  \BibitemOpen
  \bibfield  {author} {\bibinfo {author} {\bibfnamefont {J.}~\bibnamefont
  {Taylor}}\ and\ \bibinfo {author} {\bibfnamefont {M.}~\bibnamefont {Lukin}},\
  }\href@noop {} {\bibfield  {journal} {\bibinfo  {journal} {Quantum
  information processing}\ }\textbf {\bibinfo {volume} {5}},\ \bibinfo {pages}
  {503} (\bibinfo {year} {2006})}\BibitemShut {NoStop}%
\bibitem {Pedersen200747}%
  \BibitemOpen
  \bibfield  {author} {\bibinfo {author} {\bibfnamefont {L.~H.}\ \bibnamefont
  {Pedersen}}, \bibinfo {author} {\bibfnamefont {N.~M.}\ \bibnamefont
  {M{\o}ller}}, \ and\ \bibinfo {author} {\bibfnamefont {K.}~\bibnamefont
  {M{\o}lmer}},\ }\href {\doibase 10.1016/j.physleta.2007.02.069} {\bibfield
  {journal} {\bibinfo  {journal} {Physics Letters A}\ }\textbf {\bibinfo
  {volume} {367}},\ \bibinfo {pages} {47 } (\bibinfo {year}
  {2007})}\BibitemShut {NoStop}%
\bibitem {de_Lange2010}%
  \BibitemOpen
  \bibfield  {author} {\bibinfo {author} {\bibfnamefont {G.}~\bibnamefont
  {de~Lange}}, \bibinfo {author} {\bibfnamefont {Z.~H.}\ \bibnamefont {Wang}},
  \bibinfo {author} {\bibfnamefont {D.}~\bibnamefont {Ristè}}, \bibinfo
  {author} {\bibfnamefont {V.~V.}\ \bibnamefont {Dobrovitski}}, \ and\ \bibinfo
  {author} {\bibfnamefont {R.}~\bibnamefont {Hanson}},\ }\href {\doibase
  10.1126/science.1192739} {\bibfield  {journal} {\bibinfo  {journal}
  {Science}\ }\textbf {\bibinfo {volume} {330}},\ \bibinfo {pages} {60}
  (\bibinfo {year} {2010})}\BibitemShut {NoStop}%
\bibitem {CarrPurcell}%
  \BibitemOpen
  \bibfield  {author} {\bibinfo {author} {\bibfnamefont {H.~Y.}\ \bibnamefont
  {Carr}}\ and\ \bibinfo {author} {\bibfnamefont {E.~M.}\ \bibnamefont
  {Purcell}},\ }\href {\doibase 10.1103/PhysRev.94.630} {\bibfield  {journal}
  {\bibinfo  {journal} {Phys. Rev.}\ }\textbf {\bibinfo {volume} {94}},\
  \bibinfo {pages} {630} (\bibinfo {year} {1954})}\BibitemShut {NoStop}%
\bibitem {Meiboom_Gill}%
  \BibitemOpen
  \bibfield  {author} {\bibinfo {author} {\bibfnamefont {S.}~\bibnamefont
  {Meiboom}}\ and\ \bibinfo {author} {\bibfnamefont {D.}~\bibnamefont {Gill}},\
  }\href {\doibase 10.1063/1.1716296} {\bibfield  {journal} {\bibinfo
  {journal} {Review of Scientific Instruments}\ }\textbf {\bibinfo {volume}
  {29}},\ \bibinfo {pages} {688} (\bibinfo {year} {1958})}\BibitemShut
  {NoStop}%
\end{thebibliography}
\end{document}